\documentclass[journal,]{IEEEtran}
\ifCLASSINFOpdf
\else
\fi
\usepackage{multirow}
\usepackage{amsmath}
\usepackage{graphicx}
\usepackage{epstopdf}
\usepackage{bm}
\usepackage{amsfonts}
\usepackage{subfigure}
\usepackage{amssymb}
\usepackage[colorlinks,linkcolor=black,anchorcolor=black,citecolor=black]{hyperref}
\usepackage{lipsum}
\usepackage{multirow}
\usepackage[table,xcdraw]{xcolor}
\usepackage{booktabs}
\usepackage{graphics}
\usepackage{float}
\usepackage{cite}
  
\begin{document}
%
\title{CSI Sensing and Feedback: A Semi-Supervised Learning Approach}
%
%
%

\author{Haozhen~Li,
        Boyuan~Zhang,
        Xin~Liang,
        Haoran~Chang,\\
        ~Xinyu~Gu,~\IEEEmembership{Member,~IEEE}, 
        and~Lin~Zhang,~\IEEEmembership{Member,~IEEE}
\thanks{This work has been submitted to the IEEE for possible publication. Copyright may be transferred without notice, after which this version may no longer be accessible.}
\thanks{H. Li, B. Zhang, X. Liang, H. Chang, and X. Gu are with the School of Artificial Intelligence, Beijing University of Posts and Telecommunications, Beijing,
100876, China. (e-mail: \{lihaozhen, zhangboyuan, liangxin, changhaoran, guxinyu\}@bupt.edu.cn)}
\thanks{L. Zhang is with Beijing Information Science and Technology University, Beijing 100192, China. He is also with Beijing University of Posts and Telecommunications, Beijing,
100876, China. (e-mail: \{zhanglin\}@bupt.edu.cn)}
}

\maketitle

\begin{abstract}
Deep learning-based (DL-based) channel state information (CSI) feedback for a Massive multiple-input multiple-output (MIMO) system has proved to be a creative and efficient application. However, the existing systems ignored the wireless channel environment variation sensing, e.g., indoor and outdoor scenarios. Moreover, systems training requires excess pre-labeled CSI data, which is often unavailable. In this letter, to address these issues, we first exploit the rationality of introducing semi-supervised learning on CSI feedback, then one semi-supervised CSI sensing and feedback Network ($S^2$CsiNet) with three classifiers comparisons is proposed. Experiment shows that $S^2$CsiNet primarily improves the feasibility of the DL-based CSI feedback system by \textbf{\textit{indoor}} and \textbf{\textit{outdoor}} environment sensing and at most 96.2\% labeled dataset decreasing and secondarily boost the system performance by data distillation and latent information mining.
\end{abstract}


%
\IEEEpeerreviewmaketitle

\section{Introduction}
%
%
%
%
\IEEEPARstart{T}{he} downlink Channel statement information (CSI) of the multiple-input multiple-output (MIMO) plays a critical role in precoding, beamforming, and power allocation to achieve high beamforming gain through air interface. In a frequency division duplexing (FDD) communication system, downlink CSI has less reciprocity than it in time division duplexing (TDD), which leads to reporting CSI to Base station (BS) through the feedback link. However, with larger-scale antenna arrays deployed in the MIMO communication system \cite{MIMO}, feedback transmission is hindered by excessive overhead.

Recently sprung up many deep learning-based (DL-based) CSI feedback applications. Compared with the traditional method, the neural network (NN) can achieve better performance and reduce the feedback overhead \cite{2020Artificial}. A novel CSI sensing and recovery mechanism called CsiNet effectively learns from the CSI training samples proposed in \cite{CsiNet}. CsiNet+ further improved the NN structure and involved a quantization module in obtaining better performance gain \cite{Multi-rate}. Hereafter, problems such as denoise \cite{AnciNet}, lighten model complexity \cite{Lightweight} are being discussed.

The existing literature using multi-steam Autoencoders (AE) requires excess pre-labeled training data, which is significantly labor-intensive and time-consuming. In addition, practical wireless transceivers distributed layout requires a decision-making mechanism on both sides to determine which pretrained model to use. Furthermore, supervised CSI data labels are assigned arbitrarily, that latent representation of channel is ignored during model training. It is also interesting that a ``misclassified problem'' of CSI data is found when semi-supervised learning is conducted.

Semi-supervised learning is a branch of machine learning concerned with using pre-labeled and unlabeled data to perform specific learning tasks simultaneously \cite{SSsurvey}, has been proved to be one effective deep learning method in the wireless communication network \cite{SSCM1}, \cite{SSCM2}.

\begin{figure*}[htbp]
    \centering
        \includegraphics[width=6.8 in]{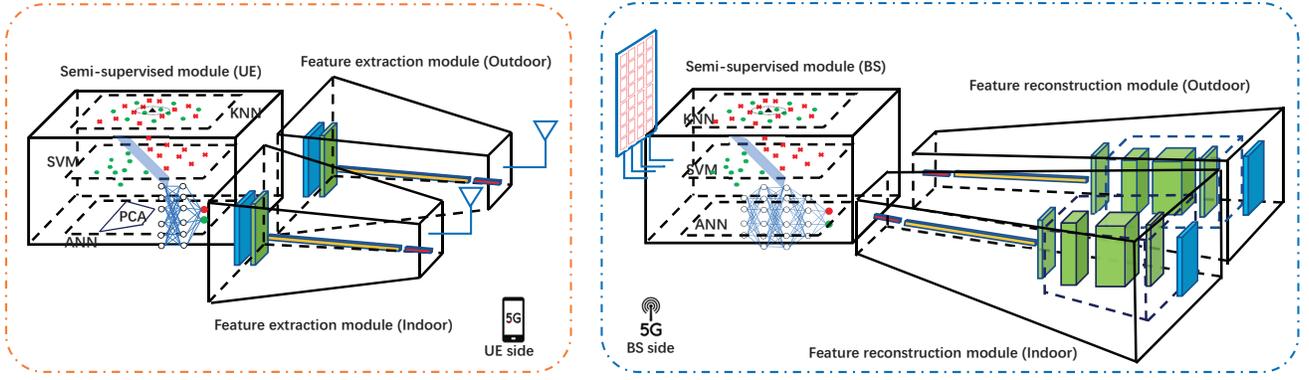}
        \caption{The architecture of $S^2CsiNet$, including a Semi-supervised module and Feature extraction encoder on the UE side, similarly a Semi-supervised module and Feature reconstruction decoder on the BS side.  }
    \label{Arch}
\end{figure*}

In this letter, we propose a semi-supervised learning CSI sensing and feedback system. In doing so, we first explore the rationality of using indoor and outdoor so-called wireless channel environments as the pseudo labels for CSI data to achieve the premise of semi-supervised learning. Later, the CSI data ``misclassified problem'' is studied. Specifically, we propose the novel scheme, $S^2$CsiNet, shown in Fig. \ref{Arch}, where the distributed network fits the layout of transceivers can handle the wireless channel environment changing. Three classifiers are compared in semi-supervised modules to leverage unlabeled data prediction to obtain additional information. One self-learning training flow for $S^2$CsiNet is introduced to support semi-supervised learning and solve the sizeable pre-labeled data problem.

To our best knowledge, $S^2$CsiNet is the first framework introducing semi-supervised learning to the DL-based CSI feedback system. Evaluation on benchmark CSI feedback networks shows that introducing semi-supervised learning is a proper path to confront the common challenges of AI in the wireless communication system.

The remainder of this paper is organized as follows. Section II shows the system model and problem statement. In section III, the proposed $S^2$CsiNet structure, network formulation, and training flow are introduced in detail. The  $S^2$CsiNet is evaluated with limitations on supervised data and different comparisons in Section IV, and section V concludes the paper.

\section{System model and Rational of Approach}
\subsection{System model}
In FDD massive MIMO system, we examine a single-cell massive MIMO transceiver pair, where the BS has $N_t$ transmit antennas, and UE has a single receive antenna. Orthogonal frequency division multiplexing (OFDM) with $N_c$ subcarriers is deployed. The downlink signal received on the UE side at the $n_{th}$ subcarrier can be expressed as $y_n=\ w_nx_n+z_n$, where $x_n$, $z_n\in\mathbb{C}$  are the transmit symbols and additional noise of wireless channel. Channel coefficient is denoted as $ w_n\triangleq h_n^Hv_n $, where $h_n\in\mathbb{C}^{N_t\times1}$ is the downlink channel vector, and $ v_n\in\mathbb{C}^{N_t\times1}$  is the precoding vector of the $n_{th}$ subcarrier, respectively.

From the original domain of CSI data, i.e., spatial-frequency domain, the downlink channel matrix is denoted by $\textbf{H}=\left[h_1\cdots h_{N_c}\right]^H\in\mathbb{C}^{N_c\times N_t}$, where $ \left(\cdot\right)^H $ expresses conjugate transpose. The critical metric transmitter precoding vector $ {v_n}'$ can be obtained from downlink channel matrix $\textbf{H}$ at the BS side. CSI sparse representation is the basic assumption of compressing and restructure process, CSI data sparsity achieved by 2-D discrete Fourier transform (DFT) from original domain to angular-delay domain denotes as
\begin{equation}
\textbf{H}=F_d\hat{H}F_a^H, 
\end{equation}
\noindent where $\bm{F}_{d} \in \mathbb{C}^{\tilde{N}_{c} \times \tilde{N}_{c}}$ and $\bm{F}_{a} \in \mathbb{C}^{N_{t} \times N_{t}}$  are DFT matrices. For channel vectors are restricted on delay domain, only the first ${\hat{N}}_c $ rows have non-zero value. Therefore, downlink CSI data parameters are reduced to $ 2{\hat{N}}_cN_t$.

With the sparsity of CSI data after 2-D DFT, AE is deployed motived by DL-based image compressing implements. One feature extraction encoder modeled as formula
\begin{equation}
\textbf{s}\ = f_{en}\left(\textbf{H}\right),
\end{equation}
to compress the $2{\hat{N}}_cN_t$ parameters $\textbf{H}$ to a given length codeword data $\textbf{s}$, with compression ratio $\gamma= 2{\hat{N}}_cN_t/N$. Feature reconstruction decoder modeled as 
\begin{equation}
\hat{\textbf{H}} = f_{de}(\textbf{s}),
\end{equation}
from codeword $\textbf{s}$ to original $2{\hat{N}}_cN_t$ parameter CSI data.

\begin{figure}[hbtp]
    \centering
    \subfigure[CSI data 3D scatter]{\includegraphics[width=3.5 in]{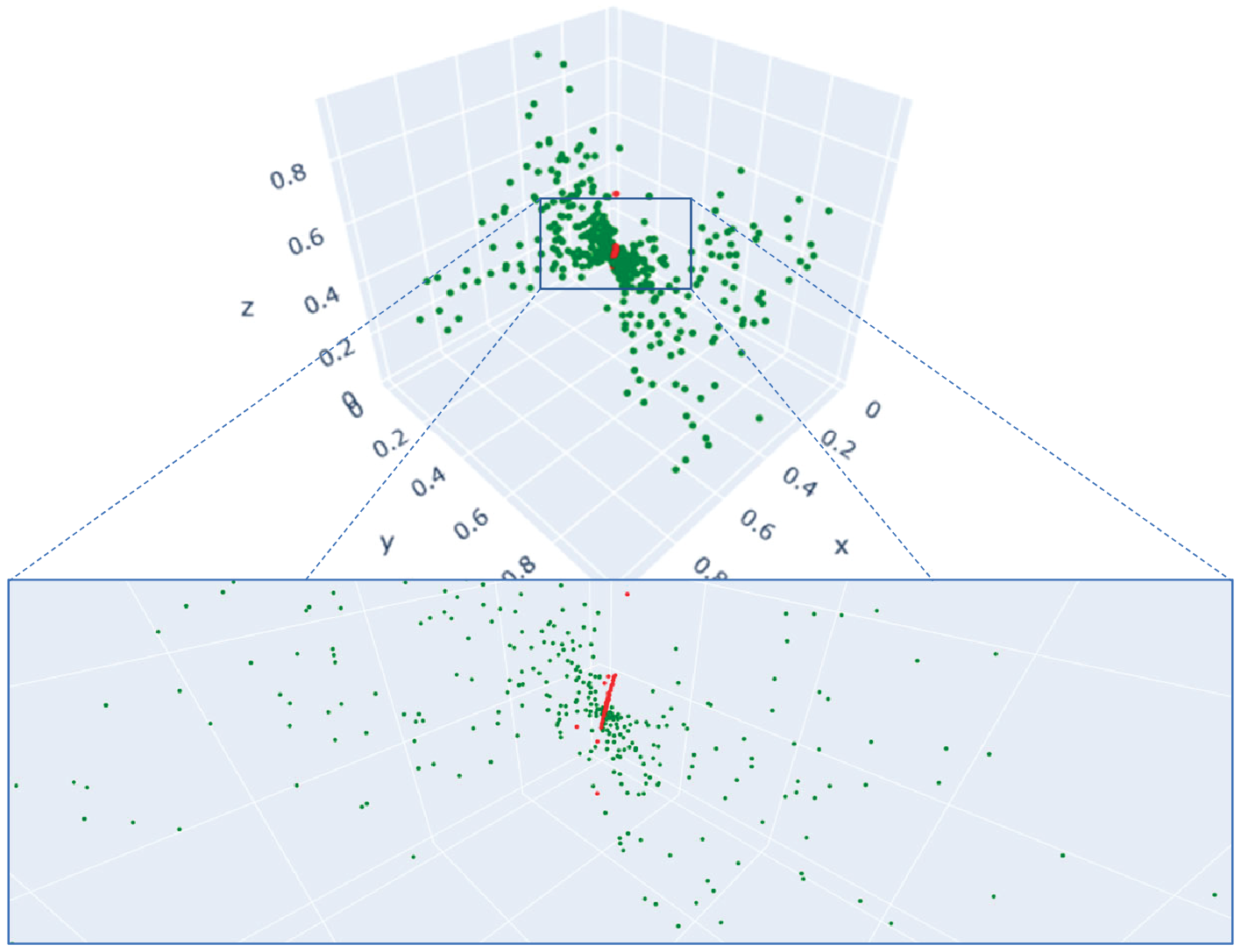}
    }
    \subfigure[CSI 2D comparison between random shuffled and classified CSI sample.]{\includegraphics[width =3.5 in]{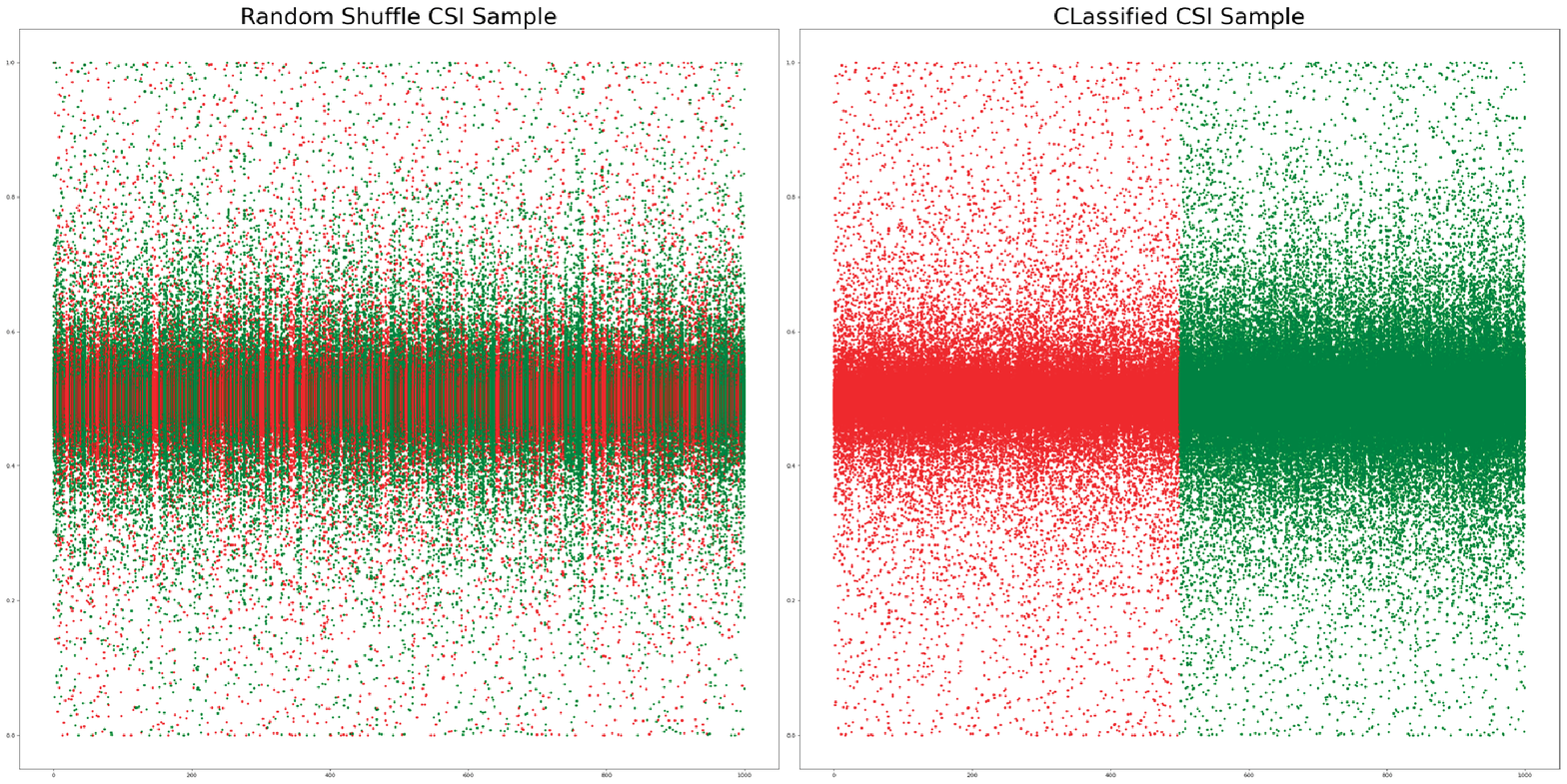}}
        
\DeclareGraphicsExtensions.
\caption{Continuity assumption of CSI data where red points from Indoor scenario and green from Outdoor.}
\label{Vis}
\end{figure}

\subsection{Problem Statement}

CSI data is a self-representation of the wireless channel environment. Considering introducing semi-supervised learning to the CSI feedback system, \cite{CSI-location} attempts to relate collocated CSI data to user location with a machine learning method, which shows the separability of CSI data.

In order to make any use of collected CSI data, a typical classifier can tap latent wireless channel environment information. CSI satisfies the continuity assumption, which is the premise of semi-supervised learning, as illustrated in Fig. \ref{Vis}. Indoor and outdoor CSI datasets show a preference for decision boundary in the low-density region that data points close to each other are more likely to share a label, and so few points are close to each other but in different classes.

\begin{figure}[hbtp]
    \centering
    \subfigure[Typical Indoor and Outdoor CSI samples]{\includegraphics[width=3.5 in]{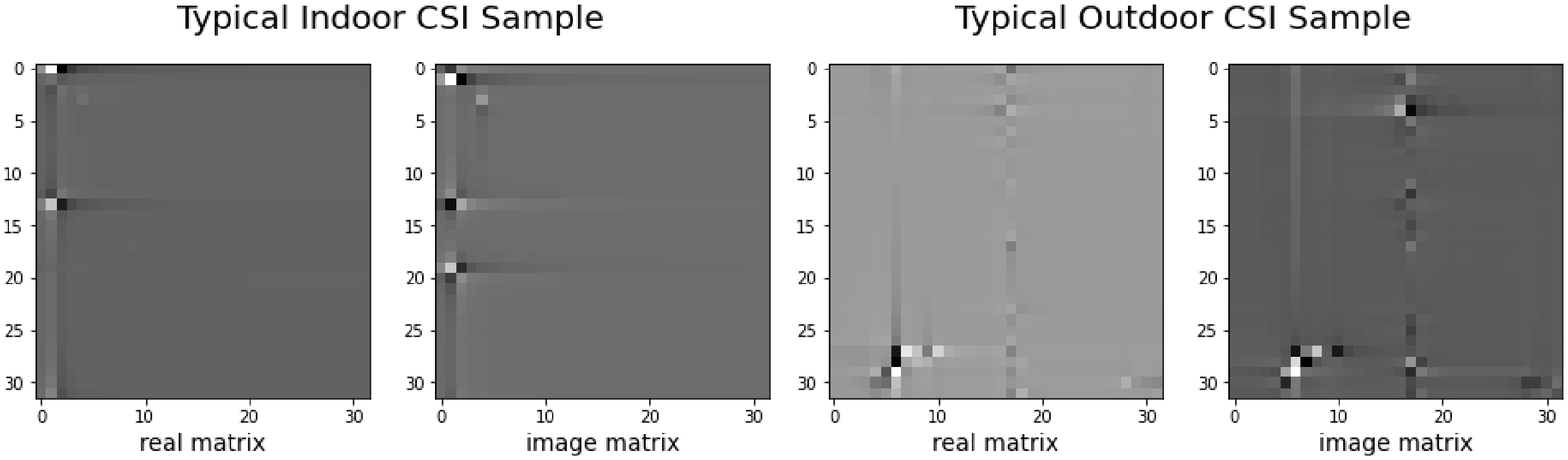}}
    \subfigure[Pre-labeled outdoor classified as indoor, and pre-labeled indoor classified as outdoor CSI samples]{\includegraphics[width =3.5 in]{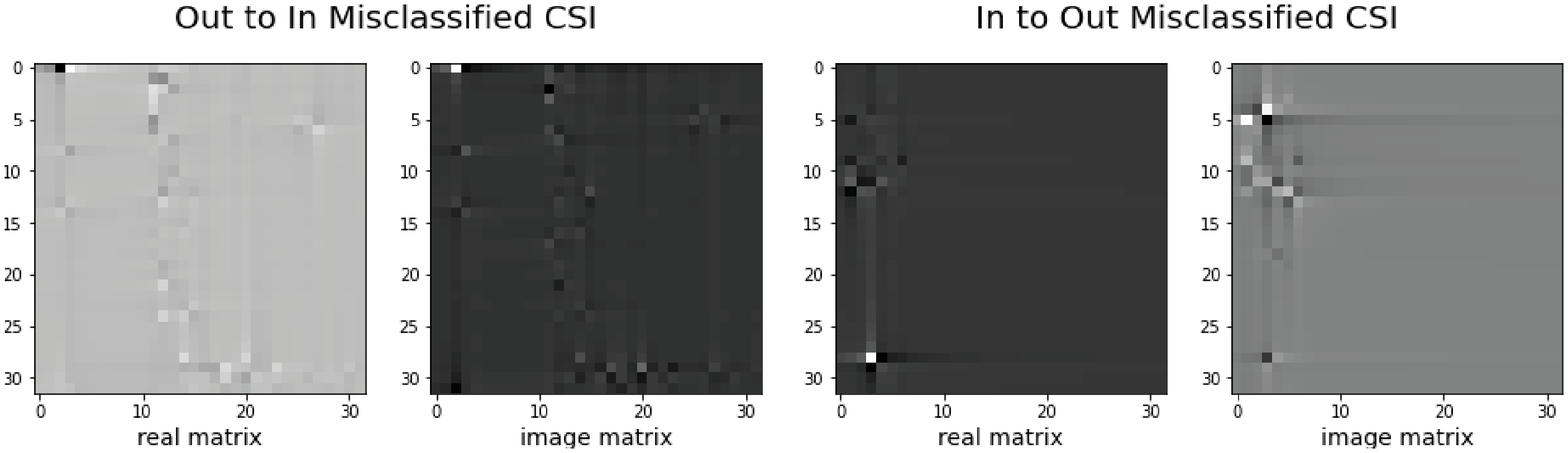}}
    
    \caption{Visualization of CSI data `` problem''}
    
    \label{Mis}
\end{figure}

Significantly, so-called CSI data ``misclassified problem'' is found when semi-supervised learning is applied. Misclassification arises even in a well-trained classifier. Visualization of CSI data misclassification shown in Fig. \ref{Mis}. The analysis of CSI misclassification is as follows. The supervised CSI dataset has a ground-true label assigned arbitrarily from the simulation dataset generator setting in \cite{COST-2100}. Due to the random variation of the wireless channel, an indoor labeled CSI may show a closer representation to an outdoor one and vice versa. Note that CSI data after a semi-supervised module favors a separation between classes.  Therefore, classified CSI datasets will be distilled as the fountain of later AE reconstruction performance gain.

For codeword sensing at the BS side, a well-trained encoder with an overfitting tendency tends to generate far different heterogeneous codewords. Therefore, a misclassified CSI will most likely lead to misclassified codeword as a ``misclassified problem'' transduction.

\section{ $S^2$CsiNet and Self-learning tranining}
\subsection{Semi-supervised module}
Semi-supervised modules in $S^2CsiNet$ have two critical tasks, primarily sensing wireless channel environment and secondarily distilling processing data. Three classification schemes, i.e., $k$-NN ($k$-nearest neighbors), SVM (Support Vector Machine), and modified ANN (Artificial Neural Network) will be adopted.

\subsubsection{$k$-NN based semi-supervised module}
In a $k$-NN classifier, CSI data will be observed from a low-dimensional subspace. The principle behind k-NN learning is to find a predefined constant as the number of training samples closest in the distance to the new point. Here we use the standard Euclidean distance.
$d_{ij} =\Vert \ x_{i}-x_{j} \Vert^2.$ 
Where $\Vert\cdot\Vert^2 $ denotes the Euclidean norm, $i,j\in R^{2048}$. As a rule-based classification method, $k$ is the crucial parameter, which is highly data-dependent. Lager $k$ suppresses the effect of noise but makes the classification boundaries less distinct. With the classifier training process, $k$ was found out with a range of integers. The probability for a test sample to obtain the label $j$ is
\begin{equation}
\bm \ P\left(y=j\middle| X=x\right)=\frac{1}{k}\sum_{n \in N} I\left(y^{(i)}=j\right),
\end{equation}
 where $\ N\in R_{KNN}^{N_{Label}}$ is the size of pre-labeled CSI training datasets.

\subsubsection{SVM based semi-supervised module}
SVM introduces kernel tricks that enable us to observe data from higher-dimensional space. One hyperplane will be found to classify labeled training samples. Given labeled training vectors $x_j\in R_{SVM}^{N_{Label}},\ i=1,\cdots,N_{Label}$ from two classes, and a vector $y\in\left\{1,-1\right\}^{\in R_{SVM}^{N_{Label}}}$. With optimal hyperparameters $\omega,b \in R$. The predication  is given by $sign\left(\omega^T\varphi\left(x\right)+b\right)$ for each sample $x$. we obtain the SVM classifier by solving the following primal problem:
\begin{equation}
\bm \min_{\omega,b,\zeta} {\frac{1}{2}\omega^T\omega}+C\sum_{i=1}^{n}\zeta_i, subject\ to\ y^T\alpha=0,
\end{equation}
\begin{equation}
\bm where\ 0\le\alpha_i\le\ C,\ i=1,\cdots,n.
\end{equation}

SVM tries to maximize margin by minimizing $\Vert\omega\Vert^2 = \omega^T\omega$, where $\zeta_i$ indicates the redundancy of misclassification, $C$ controls the strength of this penalty. The hyper-parameter $C$ and choice of kernel type will be optimal with a multi-fold test. Here we set $C$ as 10 and RBF kernel which can be approximated as $ \langle \varphi_x,\varphi_x'\rangle = K(x,x')$ , where $\varphi$ is the implicit mapping of RBF.

\subsubsection{ANN based semi-supervised module}
\begin{figure}[hbtp]
    \centering
    \includegraphics[width=3.5 in]{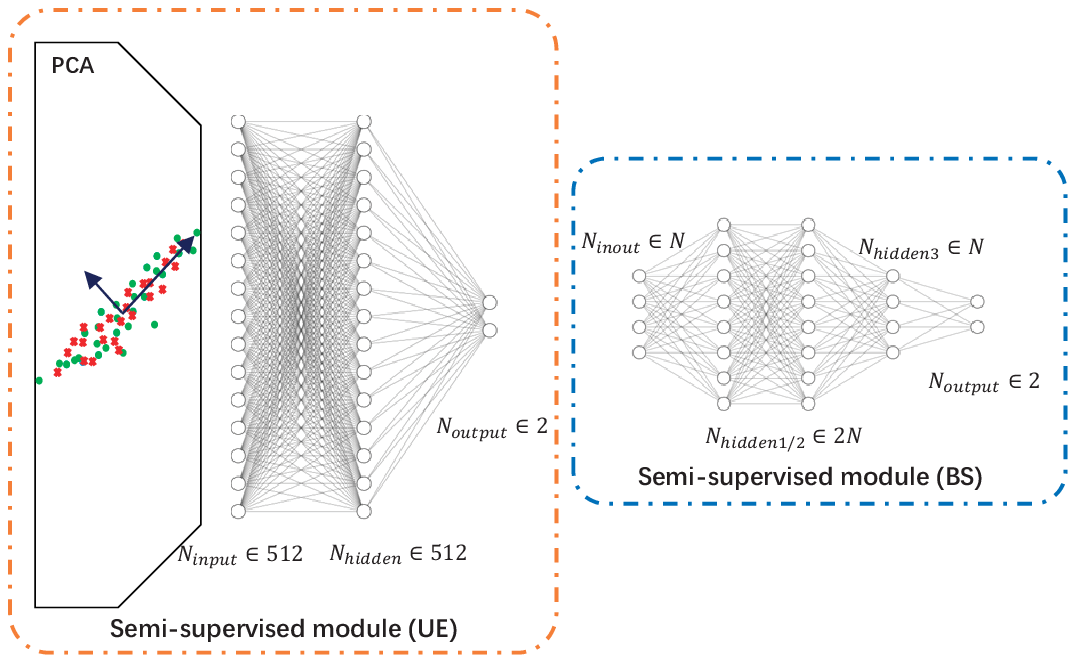}
     \caption{Illustration of the ANN architecture on the UE and the BS sides }
    \label{ANNArch}
\end{figure}

In this paper, a adopted ANN shown in Fig. \ref{ANNArch} is used at the UE side Semi-supervised module. One Principal Component Analysis (PCA) is used to compress the dimensions of the limited labeled CSI dataset to fit the empirically setting $\{ 512,512,2\}$ Multi-layer perceptron. However, the codeword processed by the BS side Semi-supervised module is low dimensional and concise. Therefore, PCA ahead ANN with $\{ N,2N,2N,N,2\}$ can be omitted.

 \begin{figure*}[hbtp]
    \centering
        \includegraphics[width=6.8 in]{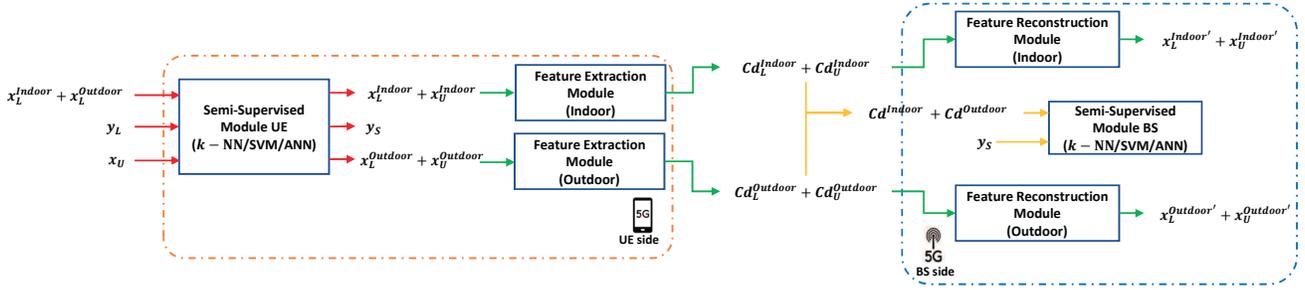}
        \caption{The illustration of $S^2$CsiNet where blue and red arrows belong to the self-learning process, green arrows belong to AE end-to-end training, yellow arrows belong to BS semi-supervised module training.
}
    \label{Flow}
\end{figure*}

\subsection{Feature Extraction and Reconstruction module}
$S^2$CsiNet can be coupled with any state-of-art CSI feedback network, for the sake of fairness in the comparison of experimental results, the CSI data feature extraction and reconstruction module of $S^2CsiNet$ are transferred from CsiNet. A two channels convolutional layer with a $3\times3$ size filter was utilized on the feature extraction module. Flatten layer appends after the convolutional layer to spread extracted features to a vector, fully connected (FC) layer with M neurons, is used to compress vector to achieve required compression ratio. The generated codewords transmit to BS side through feedback link.

An FC layer is firstly used to recover the original vector dimension from the codeword at the feature reconstruction phase. A residual block with three multi-channel convolution filters will be conducted twice, a two channels convolutional layer is stacked with a residual block to reconstruct the CSI data to a $2\times\ 32\times32$ tensor value. Mean square error (MSE) is used as the loss function of the NN, defined as follow: 
\begin{equation}
\bm L\left(\textbf{H},\hat{\textbf{H}}\right) = \frac{1}{N} \sum_{i=1}^{N}\Vert \hat{\textbf{H}}_i-\textbf{H}_i \Vert^2_2
\end{equation}

\subsection{$S^2$CsiNet Self-training Flow}

Self-learning is regarded as a compact way of training the network in a semi-supervised fashion. It first utilizes limited labeled CSI data to train the semi-supervised module \cite{SSsurvey}, then uses the well-trained model to predict the UE side collected unlabeled CSI data. Collection of classified CSI data later used to train the unsupervised two streams AE in an end-to-end way shown at the UE side of Fig. \ref{Flow}.

For convergence of classifiers, different limitations of labeled supervised data will be chosen from 70,000, 50,000, 10,000 balanced training samples prospectively. Rule-based classifier $k$-NN and model-oriented SVM and ANN will be trained to follow their loss functions with multi-folds selected optimal hyperparameters.

After the UE side semi-supervised module is well-trained, two streams of AEs for indoor and outdoor respectively are trained with classified indoor and outdoor CSI data on their loss function. For the downstream of the $S^2$CsiNet training flow, indoor and outdoor supervised and unsupervised codewords generated from the trained feature extraction module will be the input training set of the semi-supervised module on the BS side of Fig.  \ref{Flow}. With the proposed training flow, well-trained models can be fitted to the distributed layout of the wireless communication system.

\section{Performance Evaluation}
In this section, the evaluation dataset and process are described. Proposed $S^2$CsiNet uses the same CSI dataset generated in \cite{CsiNet} to compare with the CsiNet scheme fairly. This dataset consists of 100,000 training, 30,000 validation, and 20,000 test samples for indoor and outdoor, respectively. Wireless channel environment of indoor comes from default parameter setting of the indoor picocellular scenario at the 5.3 GHz band, outdoor from the semi-urban scenario at the 300 MHz band.  $N_t=32$ Uniform Linear Array (ULA) antennas at the BS-side and $N_c=1024$ subcarriers are deployed.

To simulate the practical mobile wireless communication downlink CSI data, randomly shuffled indoor and outdoor 40,000 test datasets are used to conduct the performance compared with several benchmark performances. We adopt the NMSE as the matrix of CSI reconstruction accuracy, The expression of NMSE is formulated as follows:

\begin{equation}
  NMSE = \mathbb{E}\{ \dfrac{\Vert \textbf{H} - \hat{\textbf{H}} \Vert^2_2}{\Vert\textbf{H} \Vert^2_2} \}
\end{equation}

\subsection{Distillation performance}
The $S^2$CsiNet can distillate the CSI data with different wireless channel environments. The evaluation result can be compared to the wireless environment information ignored training prototype CsiNet. 
$S^2$CsiNet shows a noticeable accuracy advantage under multiple compression ratios. The $S^2$CsiNet performance fluctuates due to the difference among the three classifiers, where the convergent $S^2$CsiNet-$k$-NN has more CSI misclassification happen than the other two $S^2$CsiNet.  The evaluation comparison shows in Table \ref{main_results}.

\begin{table}[hbtp]
\caption{Distillation performance}
\label{main_results}
\centering
\normalsize
\resizebox{\columnwidth}{!}{
\begin{tabular}{ccccccccc}
\toprule
\multicolumn{9}{c}{CSI Reconstruction Accuracy (NMSE in dB)}                                                                                                                                                                                                                                                     \\ \hline
\multicolumn{1}{c|}{}                     & \multicolumn{2}{c|}{CsiNet}                                &
\multicolumn{2}{c|}{$S^2$CsiNet-$k$-NN} &
\multicolumn{2}{c|}{$S^2$CsiNet-SVM}  &
\multicolumn{2}{c}{$S^2$CsiNet-ANN}     \\ \cline{2-9} 
\multicolumn{1}{c|}{\multirow{-2}{*}{CR}} & \multicolumn{1}{c|}{Indoor} & \multicolumn{1}{c|}{Outdoor} & \multicolumn{1}{c|}{Indoor}                         & \multicolumn{1}{c|}{Outdoor}                        & \multicolumn{1}{c|}{Indoor}                         & \multicolumn{1}{c|}{Outdoor}                        & \multicolumn{1}{c|}{Indoor} & Outdoor \\ \hline
\multicolumn{1}{c|}{4}                  & \multicolumn{1}{c|}{-17.36} & \multicolumn{1}{c|}{-8.75}   & \multicolumn{1}{c|}{\cellcolor[HTML]{FFFFFF}-17.68} & \multicolumn{1}{c|}{\cellcolor[HTML]{FFFFFF}-10.29} & \multicolumn{1}{c|}{\cellcolor[HTML]{FFFFFF}-19.93} & \multicolumn{1}{c|}{\cellcolor[HTML]{FFFFFF}-10.28} & \multicolumn{1}{c|}{-19.79} & -10.62  \\ \hline
\multicolumn{1}{c|}{16}                 & \multicolumn{1}{c|}{-8.65}  & \multicolumn{1}{c|}{-4.51}   & \multicolumn{1}{c|}{\cellcolor[HTML]{FFFFFF}-9.10}   & \multicolumn{1}{c|}{\cellcolor[HTML]{FFFFFF}-5.06}  & \multicolumn{1}{c|}{\cellcolor[HTML]{FFFFFF}-9.65}  & \multicolumn{1}{c|}{\cellcolor[HTML]{FFFFFF}-4.91}  & \multicolumn{1}{c|}{-9.79}  & -4.99   \\ \hline
\multicolumn{1}{c|}{32}                 & \multicolumn{1}{c|}{-6.24}  & \multicolumn{1}{c|}{-2.81}   & \multicolumn{1}{c|}{-7.89}                          & \multicolumn{1}{c|}{-3.15}                          & \multicolumn{1}{c|}{\cellcolor[HTML]{FFFFFF}-7.71}  & \multicolumn{1}{c|}{\cellcolor[HTML]{FFFFFF}-2.99}  & \multicolumn{1}{c|}{-8.04}  & -3.02   \\ \hline
\multicolumn{1}{c|}{64}                 & \multicolumn{1}{c|}{-5.84}  & \multicolumn{1}{c|}{-1.93}   & \multicolumn{1}{c|}{-5.60}                           & \multicolumn{1}{c|}{-2.08}                          & \multicolumn{1}{c|}{-5.63}                          & \multicolumn{1}{c|}{-1.90}                           & \multicolumn{1}{c|}{-5.61}  & -2.00      \\ \hline
\end{tabular}
}
\end{table}

The data distillation function of the semi-supervised module makes the training datasets of AE purer. Classified CSI datasets better fit the tendency of AE overfitting. With compression ratio 4, 16, 32, proposed $S^2$CsiNet-$k$-NN, $S^2$CsiNet-SVM, $S^2$CsiNet-ANN have performances overall better than the original CsiNet result shows in \cite{CsiNet}.  Under the severe compression ratio 64, the three proposed networks have a few improvements to the outdoor at the cost of very little deterioration to the indoor CSI data.

\subsection{Network capability and Pre-labeled data decreasing}

In the practical wireless communication system, wireless channels can be unpredictable changing. For proposed $S^2$CsiNet has wireless sensing ability, the Network capability can be compared with the state-of-the-art comparable evaluation method from benchmark DL-based CSI feedback system \cite{Multi-rate}, which investigated whether a neural network can handle two different scenarios simultaneously. For the sake of comparison fairness, the model preparation phase of CsiNet+ can be transferred to CsiNet. The evaluation data mixed up indoor and outdoor CSI test datasets and randomly shuffled to simulate the random variation of the wireless channel.

\begin{figure}[hbtp]
    \centering
    \includegraphics[width=3.5 in]{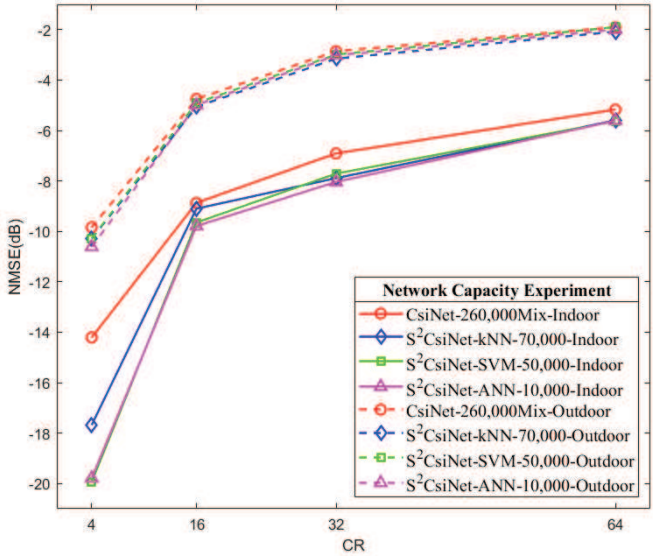}
     \caption{Network capability and pre-labeled data decreasing performance comparison.}
    \label{Netpp}
\end{figure}

Fig. \ref{Netpp} compares the NMSE performances achieved by CsiNet\cite{CsiNet}, $S^2$CsiNet-$k$-NN, $S^2$CsiNet-SVM, $S^2$CsiNet-ANN trained with 260,000, 70,000, 50,000 and 10,000 pre-labeled CSI datasets respectively. All semi-supervised approaches have comprehensive advantages compared with 260,000 pre-labeled dataset trained CsiNet under four different compression rates for indoor. It is noteworthy that under compression ratio equal to 4, the 10,000 pre-labeled ANN based semi-supervised module has nearly 6dB gain to the 260,000 pre-labeled trained unsupervised CsiNet, without any modification on AE network structure. For outdoor, all three proposed networks performances are fluctuated but basically the same as the 260,000 pre-labeled dataset trained CsiNet, which has nearly 2dB advance to 100,000 purely-pre-labeled outdoor training CsiNet performance.

Fig. \ref{Netpp} shows that, introducing semi-supervised learning can reduce pre-labeled dataset size, either in indoor or outdoor scenarios. The required supervised dataset size is dependent on the semi-supervised module classification capability and self-training strategy design. With the convergent semi-supervised module, the pre-labeled dataset size can be reduced sharply with much better reconstruction accuracy. The proposed ANN based semi-supervised module requires only 10,000 pre-labeled dataset, which is only 3.8 \% of worse performance unsupervised system.

\section{Conclusion}
In this letter, we propose a novel semi-supervised learning CSI sensing and feedback network $S^2$CsiNet. We first exploit the rationale of the approach, then formulate the $S^2$CsiNet and compare three semi-supervised modules. We show that our approach would improve the feasibility of the DL-based CSI feedback system and boost the reconstruction accuracy through numerical results. The semi-supervised approach coupling with benchmark CSI feedback system proved to be performed well at various conditions and can be used for practical scenarios. These properties make $S^2$CsiNet practical in the wireless system with time-varying channels where pre-labeled training data is unavailable.


%





\ifCLASSOPTIONcaptionsoff
  \newpage
\fi



\bibliographystyle{IEEEtran}
\bibliography{IEEEexample}

\end{document}